\documentclass[aps,pre,twocolumn,groupedaddress,showpacs]{revtex4} 
\usepackage{graphicx}  
\usepackage{subfigure}  
\usepackage{bm}  
\newcommand{\sr}{$\stackrel} 
\newcommand{\ra}{\rightarrow} 
\newcommand{\la}{\leftarrow} 
\newcommand{\ts}{\textstyle} 
\begin{document} 
 
 
\title{Impulse Backscattering in Granular Beds:\\ Introducing a Toy Model} 
 
\author{T.~R.~Krishna Mohan and Surajit Sen} 
\affiliation{Department of Physics, State University of New York, Buffalo, New York 14260-1500} 
\email{kmohan@buffalo.edu, sen@dynamics.physics.buffalo.edu} 
 
\date{\today} 
 
\begin{abstract} 
Impulses efficiently propagate into nominally dry granular beds and backscatter from buried inclusions in such beds may be potentially exploited to image shallow buried objects (SBOs). However, reliable imaging of SBOs requires ``cleaning up'' of surface vibrations, and, in addition to 3D particle dynamics simulations, a phenomenological model to parameterize the bed surface may be useful for field applications. We introduce a 1D mean-field-like toy model with two parameters that allows one to model surface vibrations, is consistent with experiments in a granular bed, and can help estimate the approximate signal transmission properties of the bed.   
\end{abstract} 
 


 
\maketitle 
 
{\it Introduction:} The imaging of shallow buried objects in a complex medium, e.g., nominally dry soil, is a difficult problem that has seen limited progress~\cite{senwdfin}. Such imaging is of relevance in connection with locating antipersonnel land mines, in archaeology, land surveying and in other applications. It has been shown that gentle mechanical impulses~\cite{rogdon} can be used to detect buried objects at depths of a meter or so in nominally dry sand beds. Detailed 3D simulations establish that nonlinear pulse propagation in 3D beds is a quasi-1D process~\cite{senetal}; normally incident pulses travel as weakly dispersive energy bundles and become more and more 1D-like with increase in area over which the impulse is generated. It would be of interest to rapidly generate images of buried backscatterers by exploiting the information contained in the time-dependent surface vibrations in granular beds~\cite{senetal}.  
 
To accomplish such imaging, it is necessary to probe some global parameter that contains coarse-grained information about grain dynamics at the bed surface. We study the space averaged, time evolution of the {\it time integrated} kinetic energies of the surface grains in idealized sand beds. Several groups have measured impulse backscattered signals at the surface in  empty beds and in beds with some buried object~\cite{rogdon}. Newtonian dynamics based 3D simulations of impulse backscattering in idealized beds have been carried out by Sen {\it et al.}~\cite{senetal}. These authors used the velocity-Verlet algorithm~\cite{velverlet} to integrate the equations of motion, and their data are consistent with the available experiments. Thus, a consensus on the spatio-temporal behavior of impulse backscattered data is beginning to emerge.   
 
It is apparent from the existing work that the energy imparted by the impulse penetrates into the system. The spread of the energy in a given $x$-$y$ plane at a given depth depends on the packing in the system. There is impulse backscattering at every granular contact. If one measures the amount of backscattered energy at the bed surface, the energy density at the bed surface rapidly depletes after the initiation of the impulse, and then rises as a function of time. The backscattering from the shallow layers is significant. The amount of backscattering from the deeper layers does get weaker, and eventually dies out. The dissipative properties of the bed play an important role in the attenuation of the impulse. 
 
We contend that, at least for the purposes of field applications, it is desirable to explore a tractable 1D toy model that can capture the critical results of the 3D simulations. In this Communication, we propose a two parameter model to describe impulse backscattering at the surface of a granular assembly. It may be necessary to use more parameters (e.g., to include information about the area across which the impulse is imparted), if one is looking for detailed agreement with experiments. The physics is similar in spirit to that of mean-field theories.  
 
\begin{table} 
\caption{\label{phase} Description of the energy transfer process in the 1D model. The arrows indicate the direction in which the layers ($lyr1$ etc.) will interact in the next time ($t$) step. The {\it exchange} case has been depicted for convenience. $e^{bs}(t)$ is the backscattered energy received back at the surface.} 
\begin{ruledtabular} 
\begin{tabular}{lccccccc} 
$t$ & $e^{bs}(t)$ &$lyr1$ & $lyr2$ & $lyr3$ & $lyr4$ & $lyr5$ & $\cdots$ \\ 
0&$0.0$&\sr{{\ts 1}}{\ra}$&$\ts 0$&$\ts 0$&$\ts 0$&$\ts 0$& $\cdots$  \\ 
1&$0.0$&\sr{{\ts (1-p)}}{\la}$&\sr{{\ts p}}{\ra}$&$\ts 0$&$\ts 0$&$\ts 0$& $\cdots$  \\ 
2&${\ts p(1-p)}$&\sr{{\ts (1-p)^2}}{\ra}$&\sr{{\ts p(1-p)}}{\la}$&\sr{{\ts p^2}}{\ra}$&$\ts 0$&$\ts 0$& $\cdots$  \\ 
3&$0.0$&\sr{{\ts p(1-p)}}{\la}$&\sr{{\ts (1-p)^2}}{\ra}$&\sr{{\ts p^2(1-p)}}{\la}$&\sr{{\ts p^3}}{\ra}$&$\ts 0$& $\cdots$  \\ 
4&$p^2(1-p)$&\sr{{\ts p(1-p)^2}}{\ra}$&\sr{{\ts p^2(1-p)}}{\la}$&\sr{{\ts (1-p)^2}}{\ra}$&\sr{{\ts p^3(1-p)}}{\la}$&\sr{{\ts p^4}}{\ra}$& $\cdots$  \\ 
\end{tabular} 
\end{ruledtabular} 
\end{table} 
{\it Phenomenological Model:} We define a vertical alignment of layers, where each layer can be thought of as a mass~\cite{th87}. At time $t=0$, we set initial energy $E=1$ for layer one and zero for the rest. At $t = 1$, the first layer in the vertical chain transfers $p$ ($<1$) of the impulse energy to the second layer, and retains $(1 - p)$. At subsequent times, the impulse will propagate in the same fashion, at every step, all the way down the chain (see Table~\ref{phase}). Each layer, after pushing the next layer in any time step, will push the preceding layer in the opposite direction in the following time step. Since the phase reverses every time step, they will interact, alternately, with layers above and below, in alternate time steps, thereby introducing significant backscattering into the problem. Note also that successive layers will be in opposite phases at any time, assuring interaction between them only in alternate time steps (Table~\ref{phase}). 
 
We model the interaction between two adjacent layers in our simulations in the following two ways: (i) {\it equipartition} case, and (ii) {\it exchange} case. In the {\it equipartition} case, the two interacting layers will come away from the interaction with equal amounts of energy; we add up the individual energies of the two layers and divide the sum equally between them. In the {\it exchange} case, we let the layers exchange their energies; the two interacting layers, after the interaction, come away with the energy of the other. The {\it equipartition} case can be viewed as one that leads to ergodic-like behavior, where we assume that the two adjacent layers get compressed to the same extent during the interaction, and the potential energy of compression gets converted back to the respective kinetic energies, which will now be half of the total energy of the two layers. The {\it equipartition} ansatz negates the symmetry breaking introduced by $p \neq 0.5$. The {\it exchange} model captures the essence of nonlinear impulse propagation in which an impulse travels as a perfect solitary wave in a 1D chain of elastic grains~\cite{Ne83}, and as a weakly dispersive energy bundle in 3D beds~\cite{senetal}; we model the situation where two energy bundles, traveling in opposite directions, go through each other without distortion. In real systems, one would expect that both the {\it equipartition} and {\it exchange} behaviors would be present, and such an extension of our study will be reported elsewhere~\cite{KM03}. 
 
We monitor the energy transfer at the surface in our model analysis. The first layer, in its negative phase (we assume positive phases to point down the chain), will transmit $p$ fraction of its energy to the surface and retain $(1 - p)$ fraction to itself; the surface does not transfer any energy back to the first layer. At the bottom of the chain, we let the last layer lose $p$ fraction of its energy in its positive phase and retain $(1 - p)$; the lost energy is presumed to travel further down in a similar fashion. These boundary conditions do not, in any way, affect our final results. We have verified our results with longer chains and there are no qualitative changes (see further discussion below); we have, therefore, employed a 40 layer long system for our studies. 

\begin{figure*} 
\centering
\begin{tabular}{cc}
\subfigure{\resizebox{3.0in}{3.0in}{\includegraphics[scale=0.5]{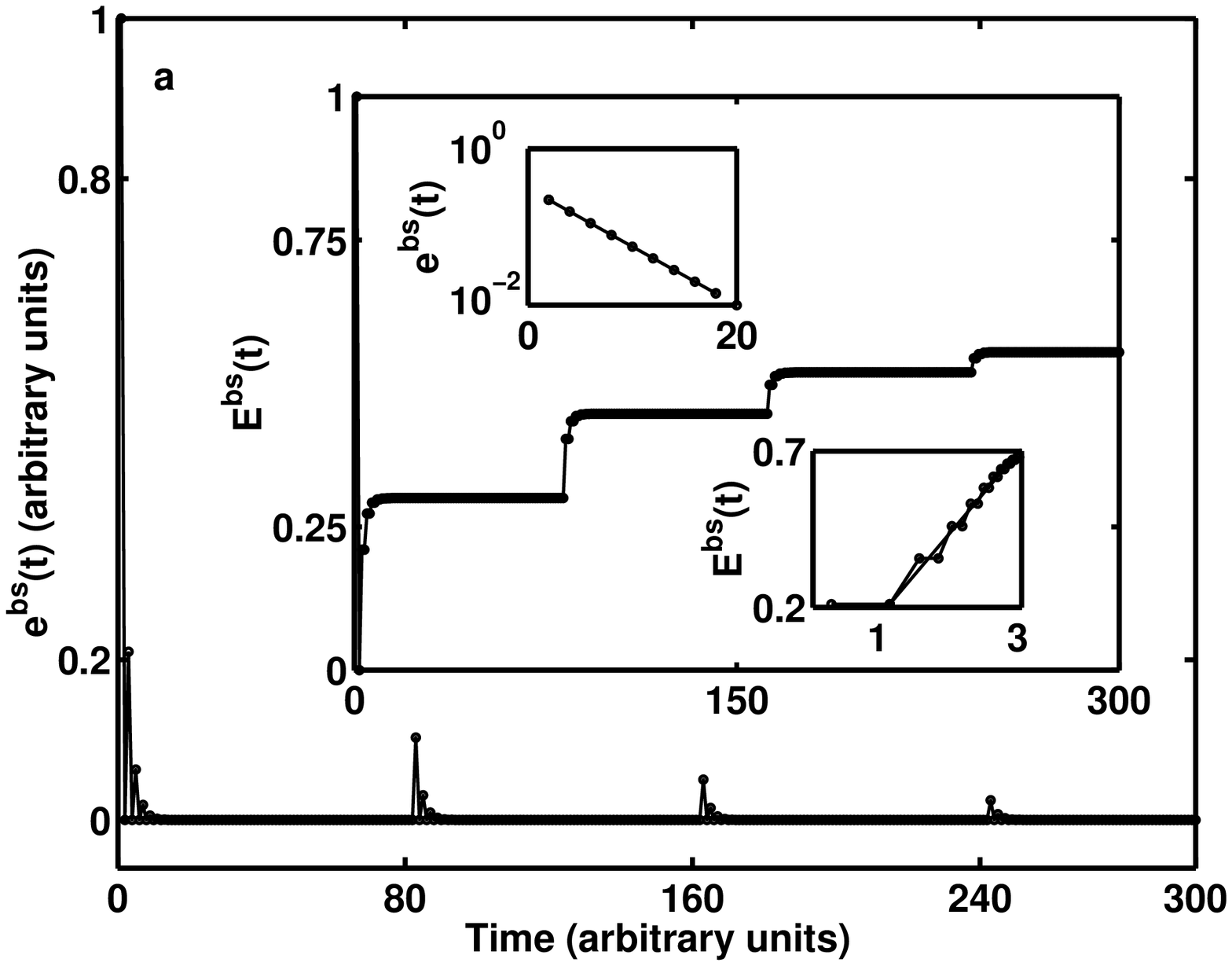}}}&
\subfigure{\resizebox{3.0in}{3.0in}{\includegraphics[scale=0.5]{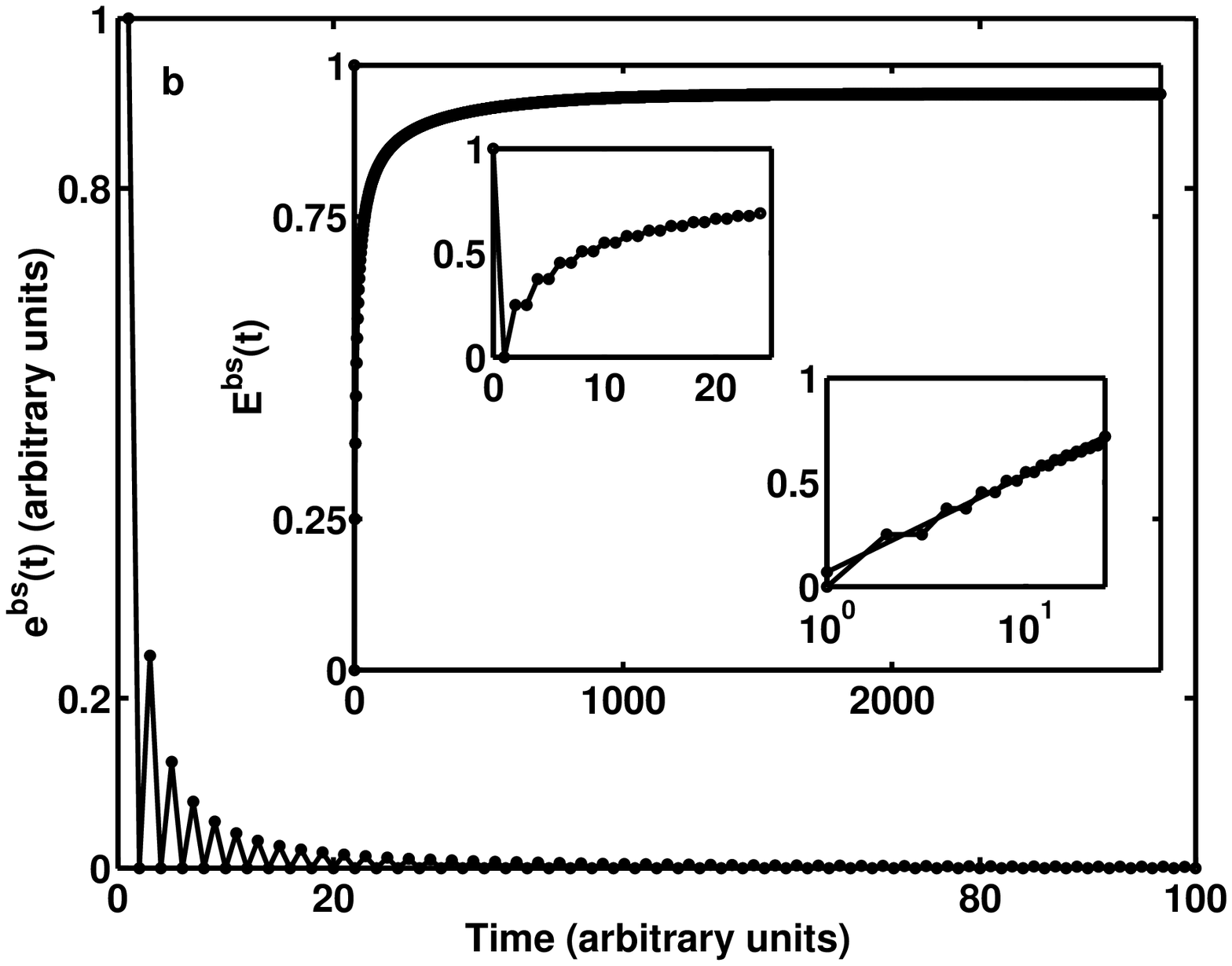}}}\\
\end{tabular}
\caption{The two cases of {\it exchange} and {\it equipartition} are shown in (a) and (b) respectively, where $p$ is constant across the layers. The two smaller insets in (a) show the exponential decay of $e^{bs}(t)$ within a subsequence (top; $y$-axis logarithmic), and the associated logarithmic growth in $E^{bs}(t)$ (bottom; $x$-axis logarithmic); the solid lines are the corresponding curve fitting lines. Smaller insets in (b) show the logarithmic growth of $E^{bs}(t)$, in the initial stages, in the {\it equipartition} case (bottom inset uses the same data as the top inset, plotted with logarithmic $x$-axis). Available 3D simulation results~\cite{senvisco} show favorable comparisons with these results.}
\label{fig1}
\end{figure*}

{\it Some Analytical results:} For the {\it exchange} model, the sequence of backscattered energy packets that arrive at the surface, $e^{bs}(t)$, can be worked out ({\it cf} Table~\ref{phase}; see Fig.~\ref{fig1}(a)):
\begin{eqnarray*}
 p(1-p), p^2(1-p), p^3(1-p), \cdots, p(1-p)^3, p^2(1-p)^3, \\ p^3(1-p)^3, \cdots, 
 p(1-p)^5, p^2(1-p)^5, p^3(1-p)^5, \cdots  
\end{eqnarray*}
and so on. Since $p < 1$, the sum of the first subsequence (each subsequence is separated by $\cdots$) yields $p$, and, similarly, the sum of the second subsequence yields $p(1-p)^2$ and so on. Thus, the entire sequence of energy packets received back at the surface can be summed to obtain, with $E^{bs}(t)$ respresenting $\sum_{t'=0}^{t'=t} e^{bs}(t')$, 
\[
E^{bs}(t=\infty)  =   p[1 + (1-p)^2 + (1-p)^4 + \cdots] = 1/(2-p) 
\]

We see that $E^{bs}(t=\infty)$ will always be $>1/2$; also, $E^{bs}(t=\infty)$ is higher if $p$ is higher. If $p < 0.5$, the successive packets of $e^{bs}(t)$ in each subsequence is going to decrease fast, and, when the next subsequence involving a higher power of $(1-p)$ arrives, a jump in $E^{bs}(t)$ takes place. In the other case, when $p > 0.5$, there is going to be a faster increase of $E^{bs}(t)$ within each subsequence itself, than brought in by the arrival of the next subsequence. The above analysis shows that a plot of $E^{bs}(t)$ against time is going to be characterized by a sequence of steps and plateaus, as is seen from Fig.~\ref{fig1}(a), first inset. The subsequences can be understood as due to the solitary wave like propagation~\cite{Ne83, senetal}, where each packet travels down the chain undiminished till it loses energy (by $p$ fraction) at the two ends. Note that the sequences are geometrically decreasing, and, as may be gathered from the plot (top smaller inset), characterized by an exponential decrease in packet sizes. As a consequence, $E^{bs}(t)$ is characterized approximately by logarithmic growth in the subsequences (bottom smaller inset), with the added feature of a saturation plateau seen in the earlier inset. 

An analysis like the above is not possible for the {\it equipartition} case because of the complicated algebra resulting from distributing half of the total energy of the two particles to each, after the interaction. Nevertheless, numerical results indicate that a similar result, of $E^{bs}(t=\infty)$ being always $>1/2$,  holds in this case as well. Also, there are no steps and plateaus in this case, except for the ones introduced by the fact that $e^{bs}(t)$ arrives in alternate time steps (Fig.~\ref{fig1}(b); see further discussion below). 

{\it Numerical Analysis:} Soil is a highly heterogeneous medium, and impulse propagation and scattering properties vary much spatially, decreasing and/or increasing with successive layers. Therefore, we investigate the above models for different scenarios characterized by different distributions of $p$. A reasonable model that can be chosen, for qualitative comparison of results for different $p$'s, is that of a logarithmic growth pattern in $E^{bs}(t)$. This is best illustrated for $p=0.25$ (for all layers) in the {\it equipartition} case (Fig.~\ref{fig1}(b)); depending on the distribution and values of $p$, the growth can be higher or lower than an average logarithmic growth in the initial stages. In Fig.~\ref{fig1}(b), it is seen that $e^{bs}(t)=0$ in alternate steps; also, the packets arrive in exponentially decreasing quanta in alternate steps. Such decay in $e^{bs}(t)$ is expected in view of the fact that most of the backscattered energy, at early times, is from the first few layers, where, by construction, the energy will be more. The deeper layers, the backscattered energy of which reaches the surface at later times, contribute to progressively higher order backscattering. The first inset to the figure shows $E^{bs}(t)$, growing rapidly in time at early times, followed by progressive slowing down. The two smaller insets in this figure shows, as a zoom-in, the region close to the origin, of the earlier inset, to show the form of the (logarithmic) growth in $E^{bs}(t)$; the bottom inset plots the data with logarithmic abscissa for explicit confirmation. This pattern has been seen in 3D simulations of impulse propagation in sand beds~\cite{senvisco}, and is consistent with experimental results~\cite{rogdon}. 
 
The maximum attained value of $E^{bs}(t)$, referred to as $E_{max}^{bs}$, is high compared to the 3D simulation results. Thus, the 1D toy model, with its restrictions in the available energy channels tends to backscatter much more energy to the surface than is typical of 3D beds. There are two ways to improve the model to better mimic the properties of 3D beds, (i) by incorporating restitution between layers, and (ii) by varying $p$ appropriately as a function of position to approximately account for changes in layer compression as a function of depth, inhomogeneities in the medium etc. We discuss the effects of these decorations in our model below. 
   
\begin{figure} 
\centering 
\includegraphics[scale=0.45]{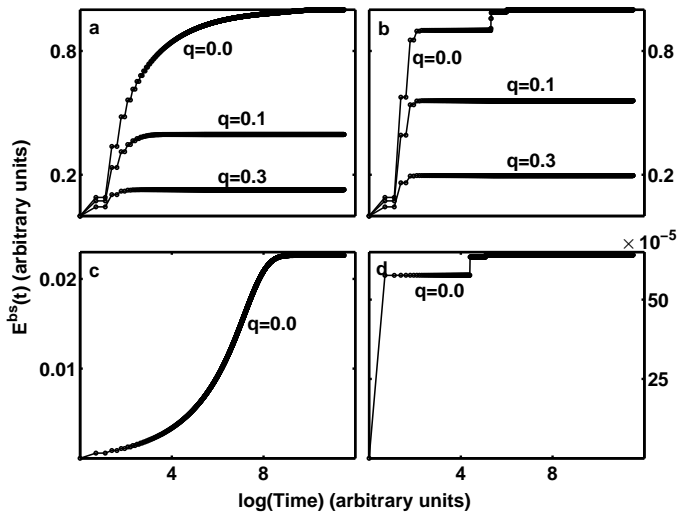} 
\caption{Growth patterns of $E^{bs}(t)$ with an inverse square law type variation in $p$. In (a) and (b), {\it equipartition} and {\it exchange} models, respectively, are shown for $p$ decreasing towards the bottom; in (c) and (d), similarly, for $p$ increasing towards the bottom. Note that the $y$-ranges in (c) and (d) are extremely small compared to those in (a) and (b). The effect of introducing a dissipation parameter, $q$, is shown in (a) and (b).} 
\label{invrs} 
\end{figure}

Introducing a tunable parameter, $q$ ($< 1$), to account for restitutional losses at each interaction (as also the spreading away of energy in 3D), which works to dissipate energy, allows us to tune $E_{max}^{bs}$; at each step, a fraction ($=q$) of the total energy of the two interacting masses is removed. It was observed (see Fig.~\ref{invrs}(a) and (b)) that $E_{max}^{bs}$ decreases exponentially with $q$; it drops by an order of magnitude as $q$ is varied from 0.1 to 0.5. 
 
Now, we consider the richness of the model by exploring various distributions of $p$ and their effect on the patterns of $e^{bs}(t)$. We will only discuss the results using the plots of $E^{bs}(t)$ which are, firstly,  easier analysed, because of the logarithmic growth pattern in the initial stages. Secondly, $E^{bs}(t)$, being an integrated quantity, is, perhaps, more suited for comparison with experimental results.  Hence, we will plot these graphs on a semi-log basis for the study, with time axis being taken logarithmically.  
 
{\it Distributions of $p$ and corresponding patterns in growth of $E^{bs}(t)$:} With a uniform distribution of $p$'s ($p$ constant across the layers), analysis of the {\it equipartition} model shows that the growth is, qualitatively, symmetric with respect to $p = 0.5$; $E_{max}^{bs}$ is the highest for $p=0.5$, and falls off when $p \neq 0.5$. Clearly, the {\it equipartition} model is working best when the initial impulse itself is transmitted down the chain with equal sharing of the energy between the leading layer and the next one in the chain. On the other hand, the curves are also symmetric, on either side of $p=0.5$, around the mid-value of the respective ranges, i.e. around $0.25$ and $0.75$. The latter symmetry is with respect to the initial growth rates. The curves are concave (lesser than average logarithmic growth rate), in the initial stages, on the semi-log plots for $p < 0.25$ ($p > 0.75$), and convex (larger than average logarithmic growth rate) for $p > 0.25$ ($p < 0.75$). This symmetry is natural because of the symmetry introduced by the exchange of roles between $p$ and $(1-p)$ as $p$ is varied. We get the best (approximate) logarithmic growth, in the initial stages, for $p=0.25$ ($p = 0.75$). 

In the {\it exchange} case, there is no such symmetry. It is a monotonous behavior with $E^{bs}(t)$ requiring many more cycles to saturate, the lower the $p$ value is, and saturating faster, the higher the $p$ value is. In the latter case, if the $p$ value is sufficiently high, $E_{max}^{bs}$ can be attained almost in the first cycle. This is understandable since the undispersed energy bundles that travel up and down the chain, but for the losses at the two ends, are smaller (larger) in size when $p$ is smaller (larger).
 
We may let $p$ decrease towards the bottom and vice versa. In the {\it equipartition} case, if we impose a linear decrease (increase), in the initial stages, the graphs are convex (concave) and the variation in the amount of decrease (increase) shows variation in convexity (concavity). This is due to the longer presence of energy in the upper layers, with higher (lower) $p$ values . Similarly, for the {\it exchange} model as well, steeper initial growth is achieved for higher $p$'s in the top layers and vice versa. We may impose other rates of decrease (increase) like exponential, inverse square law etc. but the basic charateristic of the convexity (concavity)/steepness of the {\it equipartition/exchange} graphs being dictated by the presence of larger (smaller) $p$'s in the upper layers will remain the same; only it will be more pronounced with larger rates of decrease (increase) like exponential or inverse square law etc. In Figs.~\ref{invrs}(a) and (b), we have shown results for a case where $p$'s decrease in inverse square law fashion towards the bottom for the {\it equipartition} and {\it exchange} models, respectively; note the high values of $E_{max}^{bs}$. Figs.~\ref{invrs}(c) and (d) show the results, similarly, for a case where the $p$'s increase towards the bottom; note that, if the $p$'s in the upper layers are extremely small (of the order of, say, $10^{-3}$), $E_{max}^{bs}$ is going to be extremely small ($<10^{-3}$) in both the models. 

If we let the $p$'s vary randomly, we get a combination of the effects discussed earlier and the model is sensitive to different realizations of the random distribution of $p$'s, a desirable feature for a simple phenomenological model for a highly heterogeneous medium like soil. We have also checked our results with longer chains. It is clear that the cyclicity in the {\it exchange} model will change with the length of the chain; the cyclicity is twice the length of the chain because we are only monitoring the energy at one end. This introduces, for example, a longer plateau length at each step of the staircase, in certain situations. However, $E_{max}^{bs}$ remains the same in the case of a constant $p$ across the layers. In the {\it equipartition} case, $E_{max}^{bs}$ does change slightly; however, the convexity (concavity) patterns in the plots, which depend on the $p$ values, are not changed, only they get more accentuated.  
 
In summary, it can be stated that our 1D model is robust, and, at the same time, sensitive to changes in the distribution of $p$ across the layers. The model qualitatively  reproduces the time evolution of average energy of the surface grains of  a complex, heterogeneous medium like granular beds, following the generation of a normal impulse at the bed surface. The model promises to be a useful tool for further research on this important topic.   
 
{\it Acknowledgments}:  The research has been supported by NSF CMS 0070055 and by Sandia National Labs.

\end{document}